\documentclass[final,5p,times,twocolumn]{elsarticle}

\usepackage{amsmath,amssymb,amsfonts}
\usepackage{slashed}
\usepackage{bbm}
\usepackage{nicefrac}
\usepackage{dsfont}
\usepackage{fixmath}
\usepackage{mathtools}

\usepackage{amssymb}

\newcommand{\alg}[1]{\mathfrak{#1}}
\newcommand{\grp}[1]{\mathrm{#1}}

\newcommand\ii{\textup{i}}
\newcommand\ee{\textup{e}}

\journal{Physics Letters B}

\newcommand{\de}{\operatorname{d}\!}
\newcommand{\scaladj}{\phi^{\text{adj}}}
\newcommand{\scalfun}{\phi^{\text{fun}}}
\newcommand{\bscalfun}{\hat{\phi}^{\text{fun}}}
\newcommand{\scalsing}{\phi^{\text{sing}}}
\newcommand{\Zadj}{Z^{\text{adj}}}
\newcommand{\Zfun}{Z^{\text{fun}}}
\newcommand{\sogrp}{\operatorname{SO}}
\newcommand{\sugrp}{\operatorname{SU}}
\DeclareMathOperator{\tr}{tr}

\newcommand{\bdryk}{\mathcal{K}}
\newcommand{\Rmat}{\mathcal{R}}
\newcommand{\C}{\mathbb{C}}

\newcommand{\scalq}{\tilde{\phi}}
\newcommand{\funsc}{q_{\text{cl}}}
\newcommand{\funsq}{\tilde q}

\begin{document}

\begin{frontmatter}



\title{The dilatation operator for defect conformal $\mathcal{N}=4$ SYM}

\author{Asger C. Ipsen$^a$ and Kasper E. Vardinghus$^b$}

\address[label1]{Institute of Mathematics and Institute of Physics, Humboldt-Universität zu Berlin,
\\ Zum Gro\ss{}en Windkanal 6, 12489 Berlin, Germany }

\address[label2]{Niels Bohr Institute, Copenhagen University,\\
Blegdamsvej 17, 2100 Copenhagen \O, Denmark}

\begin{abstract}
We compute the dilatation operator for local ``open string'' operators situated at the interface of a certain supersymmetric defect version of $\mathcal{N}=4$ super-Yang-Mills theory.
 This field theory is dual to a probe D5-brane intersecting a stack of D3-branes where the number of D3-branes can change between the two sides the interface.
DeWolfe and Mann obtained the dilation operator in the special case of an equal number of D3-branes.
Using a combination explicit field theory calculations and integrability considerations we 
are able to  extend this result to the general case.
\end{abstract}

\begin{keyword}
 AdS/CFT correspondence, defect CFT, integrability, probe branes, spin chains




\end{keyword}

\end{frontmatter}

\section{Introduction}

In this paper we will study a 1/2-BPS deformation of $\mathcal{N}=4$ super-Yang-Mills theory
(SYM) which is obtained
by introducing a flat defect\cite{DeWolfe:2001pq,Erdmenger:2002ex}.
We will consider the case where the rank of the gauge group jumps when crossing the defect.
To be more specific, let us put the defect at $x_3 = 0$. 
We then take the gauge group to be U$(N)$ for $x_3 > 0$, and U$(N-k)$ for $x_3 \leq 0$.
We will always be in the planar (i.e.~$N\to\infty$) limit, but keep $k$ finite.
The string dual of this setup is a single D5-brane intersecting a stack of D3s, with $k$ of the D3s coming
from the $x_3 > 0$ side dissolving in the D5\cite{Karch:2000gx}. 

Our main subject is the spectrum, at one loop, of local operators on the defect.
The spectral problem in the bulk (i.e.~away from the defect) is very well-understood;
for scalar single-trace operators, $\mathcal{O} = \tr[\phi\phi\cdots\phi]$, the one-loop
correction to the dimension is  described by a closed (i.e.~with periodic boundary conditions)
integrable spin chain\cite{Minahan:2002ve}.

On the defect, and for $k=0$, the natural analogues of the scalar single-trace operators are
$\mathcal{O} = q^\dagger \phi\phi\cdots\phi q$. The scalar defect field $q$ is the bosonic
component of a 3d hyper-multiplet\cite{DeWolfe:2001pq}, and is in the fundamental of U$(N)$. For this class of operators the one-loop spectrum 
is again controlled by an integrable spin chain\cite{DeWolfe:2004zt}, but now with 
open boundary conditions. For $k>0$ there are no $q$ fields and one should instead
consider operators $\mathcal{O} = (\bscalfun)^\dagger \phi\phi\cdots\phi \bscalfun$.
Here $\bscalfun$ is a  gauge fundamental defect field recently constructed
in Ref.~\cite{deLeeuw:2017dkd}. It is defined by a certain limit
of the bulk fields, see  \ref{app:bdry-fields}.
Our main result is the construction of an integrable spin chain for these operators
for all $k>0$, generalising the results of \cite{DeWolfe:2004zt}.

The case of $k=1$ is especially attractive from a computational point
of view; on the one hand it avoids  having to deal with the 3d hyper-multiplet
and the complicated non-linear boundary conditions for the bulk fields which
are present for $k=0$.
On the other hand it also avoids the classical scalar VEVs appearing at $k\geq 2$
which leads to a quite complicated perturbative setup\cite{Buhl-Mortensen:2016jqo}.
In Section \ref{sec:field-theory} we take advantages of this fact to directly calculate
the dilatation operator for $k=1$.

There is by now a significant body of evidence\cite{DeWolfe:2004zt,Correa:2011nz,deLeeuw:2015hxa,Buhl-Mortensen:2015gfd,Buhl-Mortensen:2017ind} 
 that the D3-D5 defect theory retains
the integrability of planar $\mathcal{N} = 4$ SYM, and
in Section \ref{sec:spin-chains} we use this as a working assumption. Generalising
recent work\cite{Pozsgay:2018dzs}, we then find a class
of integrable open spin chains, parametrised by $k$,
with symmetries matching those of our spectral problem. For $k= 1$ we reproduce
the explicit dilation operators from field theory. We thus propose to identify
these spin chains with the dilatation operator for all $k$. A direct check of this
conjecture in field theory would be desirable, but is left for the future.

Finally, in Section \ref{sec:reflection}, we calculate a certain ratio of reflection phases in 
our proposed spin chain. This ratio
is already known on the basis of the supersymmetry preserved by the defect\cite{DeWolfe:2004zt,Correa:2008av,Correa:2011nz}.
We find complete agreement for all $k$.

\section{Field theory computation}
\label{sec:field-theory}

The defect clearly breaks translation invariance along $x_3$, so there 
can be no 4d supersymmetry.
However, it turns out that 3d  $\mathcal{N} = 4$ supersymmetry is preserved.\cite{DeWolfe:2001pq}
In term of this  the bosonic degrees of freedom of the bulk
are as follows: The adjoint scalars $\phi_{1,2,3}$ together with $A_3$ form 
a hyper-multiplet, while $\phi_{4,5,6}$ and $A_{0,1,2}$ form a 
vector-multiplet.\cite{DeWolfe:2001pq}
The R-symmetry is broken by the defect as $\sogrp(6)\to \sugrp(2)_H\times \sugrp(2)_V$, 
where $\sugrp(2)_{H(V)}$ rotates $\phi_{1,2,3}$ ($\phi_{4,5,6}$).

For $k=0$ there is an additional 3d hyper-multiplet on the defect, which couples
to the bulk fields\cite{DeWolfe:2001pq}. In contrast, for
$k>0$, there are \emph{no} independent defect degrees of 
freedom\cite{Gaiotto:2008sa}.\footnote{In Appendix D of Ref.~\cite{deLeeuw:2017dkd}
  the counting of boundary operators assumes that the 3d defect hyper-multiplet is also
  present for $k>0$. This is not correct for the field theory dual to the D3-D5 setup.
  However, none of the  results in the main text of \cite{deLeeuw:2017dkd} depend
  on this counting.}
Instead, the bulk fields satisfy  specific  boundary 
conditions\cite{Gaiotto:2008sa,Mikhaylov:2014aoa}. 
This difference between $k=0$ and $k>0$ might seem unintuitive, but, as we 
illustrate in \ref{app:moving-branes}, the two cases are actually continuously connected.

\subsection{Dilatation operator for $k=1$}
In this subsection we will compute the dilatation operator for scalar open-string operators.
Since similar calculations can be found in e.g.~\cite{Minahan:2002ve,DeWolfe:2004zt,Rapcak:2015lhn}
 we will suppress some details.

For $x_3 \geq 0$ the SYM fields are $N\times N$ hermitian matrices. We decompose them as
\begin{equation}
  \phi = \begin{pmatrix}
           \scalsing &(\scalfun)^\dagger \\
           \scalfun  &\scaladj
         \end{pmatrix}\, ,
  \label{eq:block-structure}
\end{equation}
where $\scaladj$ is a $(N-1)\times (N-1)$ matrix and $\scalfun$ is a vector of length $N-1$.
The single component $\scalsing$ will not play any role in this section.
At the defect $\scaladj$ is in the adjoint of the gauge group U$(N-1)$, while $\scalfun$ is in
the fundamental.

The $\scaladj$ block joins continuously with the corresponding $(N-1)\times(N-1)$ field
living at $x_3< 0$. We normalise the propagator as (here and in the following we suppress
all colour structure)
\begin{equation}
  \langle \scaladj_i(x) \scaladj_j(y) \rangle_0 = \frac{\delta_{ij}}{|x-y|^2}\, ,
\end{equation}
with the understanding that $\scaladj_i(x) := \phi_i(x)$ for $x_3 < 0$.
The $\scalfun_i$ block satisfies Neumann (Dirichlet) boundary conditions for $i=1,2,3$
($i=4,5,6$),
\begin{equation}
  \langle (\scalfun_i(x))^\dagger \scalfun_j(y) \rangle_0
    = \delta_{ij}\left(\frac{1}{|x-y|^2} \pm_i \frac{1}{|x^R-y|^2}\right)\, ,
  \label{eq:fun-bc}
\end{equation}
with
\begin{equation}
  \pm_{i} := \begin{cases}
                +  & i = 1,2,3 \\
                -  & i = 4,5,6
              \end{cases}\, .
\end{equation}
Here the superscript $R$ denotes reflection in the defect, i.e.~inverting the sign of $x_3$.

We wish to compute the one-loop dilatation operators for scalar operators on the defect of the form
\begin{equation}
  \hat{\mathcal{O}}_{i_0,i_1,\ldots,i_{L},i_{L+1}} := (\scalfun_{i_0})^\dagger \scaladj_{i_1}
    \cdots \scaladj_{i_L} \scalfun_{i_{L+1}} 
    \leftrightarrow |\phi_{i_0}\phi_{i_1}\cdots\phi_{i_{L+1}}\rangle \, .
  \label{eq:bdry-op}
\end{equation}
Due to the boundary conditions \eqref{eq:fun-bc} this is the zero operator for 
$i_0,i_{L+1} = 4,5,6$. We thus restrict to having $i_0,i_{L+1}=1,2,3$.
Note that the operators \eqref{eq:bdry-op} are only gauge invariant on the defect; for $x_3 > 0$ the
gauge group is enhanced to U$(N)$ and \eqref{eq:bdry-op} would transform non-trivially
under this.

The dilatation operator can be read off from the renormalisation-scale dependence of 
correlation functions with an insertion of $\hat{\mathcal{O}}_{i_0,i_1,\ldots,i_{L},i_{L+1}}$
using the Callan-Symanzik equation. In fact, since we are in the planar
limit, the dilatation operator is of the nearest-neighbour form and we can thus 
treat each pair of adjacent fields in $\hat{\mathcal O}$ separately.

For a pair of adjoint scalars, the only relevant planar one-loop correction is due
to the $\tr([\phi_i,\phi_j][\phi_i,\phi_j])$ vertex (all other diagrams are proportional
to the identity in flavour space),
\begin{multline}
  \label{eq:bulk-diagram}
  \left\langle[\scaladj_i\scaladj_j](0)\scaladj_{i'}(x)\scaladj_{j'}(y)\right\rangle_{\text{1-loop, $\phi^4$}} =\\
   g^2(\delta_{ii'}\delta_{jj'}-2\delta_{ij'}\delta_{ji'}+\delta_{ij}\delta_{i'j'}) 
   \int\de z \frac{1}{|z-x|^2|z-y|^2}\left[\frac{1}{|z|^4}\right]_\mu ,
\end{multline}
where $g^2$ denote the 't-Hooft coupling up to numerical factors. 
We use $[|z|^{-4}]_\mu$ to denote the UV-renormalisation of $|z|^{-4}$. The explicit expression is
given in \ref{app:renorm}. Taking the scale derivative we find (see \ref{eq:full-space-mu-deriv}),
\begin{multline}
  \label{eq:bulk-scale-deriv}
  \mu\frac{\partial}{\partial\mu}
    \left\langle[\scaladj_i\scaladj_j](0)\scaladj_{i'}(x)\scaladj_{j'}(y)\right\rangle_{\text{1-loop, $\phi^4$}} \\
    = \frac{g^2}{8\pi^2}(\delta_{ii'}\delta_{jj'}-2\delta_{ij'}\delta_{ji'}+\delta_{ij}\delta_{i'j'})
       \left\{\frac{1}{|x|^2|y|^2}\right\} \, .
\end{multline}   
Turning to the ``left  end''
of $\hat{\mathcal{O}}$ we now need to consider 
\begin{multline}
  \left\langle[(\scalfun_i)^\dagger\scaladj_j](0)\scalfun_{i'}(x)\scaladj_{j'}(y)\right\rangle_{\text{1-loop, $\phi^4$}}\\
   = g^2(\delta_{ii'}\delta_{jj'}-2\delta_{ij'}\delta_{ji'}+\delta_{ij}\delta_{i'j'}) 
  \int\de z \left(\frac{1}{|z-x|^2}\pm_{i'}\frac{1}{|z^R-x|^2}\right) \\
  \times \frac{1}{|z-y|^2}\left[\theta(z_3)\frac{2}{|z|^4}\right]_\mu .
\end{multline}
The factor of two in the square bracket is due to the Neumann boundary conditions, 
and we have put an explicit theta function to restrict the integration domain.
The scale derivative is (see \ref{eq:half-space-mu-deriv})
\begin{multline}
  \mu\frac{\partial}{\partial\mu}
    \left\langle[(\scalfun_i)^\dagger\scaladj_j](0)\scalfun_{i'}(x)\scaladj_{j'}(y)\right\rangle_{\text{1-loop, $\phi^4$}} \\
    = \frac{g^2}{8\pi^2}(\delta^H_{ii'}\delta_{jj'}-2\delta^H_{ij'}\delta^H_{ji'}+\delta^H_{ij}\delta^H_{i'j'})
       \left\{\frac{2}{|x|^2|y|^2}\right\}\, .
  \label{eq:end-scale-deriv}
\end{multline}
Here we define $\delta^H_{ij} := \delta_{ij}(1\pm_i 1)/2$, and, for later convenience, we set 
$\delta^V_{ij} := \delta_{ij} - \delta^H_{ij}$.
The calculation for the right end is completely analogous, so we do not write it explicitly.

In \eqref{eq:bulk-scale-deriv} and \eqref{eq:end-scale-deriv} the curly bracket is the
corresponding tree-level result. By Callan-Symanzik we can thus read of 
the dilatation operator as
\begin{equation}
  H^{k=1} := \mathcal{P}^H_0\mathcal{P}^H_{L+1}
       \left(\sum_{r=0}^L [2-2P_{r,r+1} + K_{r,r+1}]\right)\mathcal{P}^H_0\mathcal{P}^H_{L+1}\, .
  \label{eq:H-k-eq-one}
\end{equation}
We use the usual spin chain language as  indicated in \eqref{eq:bdry-op}.
The subscripts indicate which sites the various operators act on, 
 $\mathcal{P}^H := \sum_{i=1}^3 |i\rangle\langle i|$ project onto the ``hyper'' subspace,
and we define
$ P|i_1, i_2\rangle := |i_2, i_1\rangle$, 
$K |i_1, i_2\rangle := \delta_{i_1 i_2}\sum_{j=1}^6 |j, j\rangle$,
as usual. The identity part of \eqref{eq:H-k-eq-one} is fixed by demanding that the 
BPS vacuum
\begin{equation}
  \hat{\mathcal{O}}^{\omega}_L := (\Zfun)^\dagger (\Zadj)^L \Zfun\, ,
  \qquad Z := \phi_1 + \ii \phi_2
\end{equation}
is annihilated. Note that the bulk part of \eqref{eq:H-k-eq-one} is identical
to the usual Hamiltonian\cite{Minahan:2002ve} for closed-string operators in SYM with no defect.
Indeed, the contribution \eqref{eq:bulk-diagram} is not affected by the defect, and this is also
the case for $k>1$.

\section{Integrable open spin chains}
\label{sec:spin-chains}

In this section we demonstrate that the dilatation operator as computed in the previous section, Eq.~\eqref{eq:H-k-eq-one}, corresponds to the Hamiltonian of an integrable open spin chain. Furthermore, the Hamiltonian is contained in a family of integrable open spin chain Hamiltonians parametrised by the size of an $\alg{su}(2)$ representation.

For $k>1$ the scalar open-string operators takes the form 
\begin{equation}
  \label{eq:bdry-op-gen-k}
   (\bscalfun_{i_0,n_0})^\dagger \scaladj_{i_1}
    \cdots \scaladj_{i_L} \bscalfun_{i_{L+1},n_{L+1}} \, .
\end{equation}
As in the $k=1$ case (Eq.~\eqref{eq:bdry-op}) we restrict $i_0,i_{L+1} = 1,2,3$, but
now the boundary fields
 have an additional SU$(2)_H$ index $n = 1,2,\ldots,k$,
see \ref{app:bdry-fields}.
We propose that the class of integrable Hamiltonians obtained in this section exactly corresponds to the
one-loop dilatation operator for operators of the form \eqref{eq:bdry-op-gen-k}. Our Hamiltonian will be
of the nearest neighbour type, 
\begin{equation}
H = \sum_{r=0}^L H_{r,r+1}\, . \label{eq:Hform}
\end{equation} 
As remarked at the end of the previous section, the bulk terms must be 
\begin{equation}
  \label{eq:Hform-bulk}
  H_{r,r+1} = 2-2P_{r,r+1} + K_{r,r+1}\, , \qquad r = 1,2,\dots,L-1
\end{equation}
in order to match field theory. Our task is thus to determined $H_{0,1}$ and $H_{L,L+1}$ with $\sugrp(2)_H\times \sugrp(2)_V$
symmetry in the correct representation, and such that the full $H$ is integrable.

An integrable spin chain with $\grp{SO}(6)$ symmetry where all sites transform in the fundamental representation can be constructed starting from the $\Rmat$-matrix \cite{Reshetikhin:1986vd}
\begin{equation}
  \Rmat_{i,j}(u) := \frac{1}{2}\left(u(u+2) + (u+2)P_{i,j} - uK_{i,j}\right)\, .
\end{equation}
Let $V_i \simeq \C^6$ for $i=1,\ldots,L$ denote the vector spaces of the bulk sites. 
The $\Rmat$-matrix is an endomorphism on the tensor 
product space $V_i\otimes V_j$, depends on the spectral parameter $u\in \C$, and satisfy the Yang-Baxter equation
\begin{equation}
\Rmat_{1,2}(u) \Rmat_{1,3}(u+v) \Rmat_{2,3}(v) = \Rmat_{2,3}(v) \Rmat_{1,3}(u+v) \Rmat_{1,2}(u)  \, . \label{eq:ybe}
\end{equation}

Given an $\Rmat$-matrix one can construct open boundary conditions for the spin chain that preserve integrability from solutions, $\bdryk$, of the reflection (or boundary Yang-Baxter) equation 
\begin{multline}
\Rmat_{1,2}(u-v) \bdryk_{1B}(u) \Rmat_{1,2}(u+v) \bdryk_{2B}(v)  \\
= \bdryk_{2B}(v) \Rmat_{1,2}(u+v) \bdryk_{1B}(u) \Rmat_{1,2}(u-v) \,  \label{eq:bybe}
\end{multline}
following Sklyanin\cite{Sklyanin:1988yz}.
To account for the additional $\grp{SU}(2)_H$ index on the boundary fields we shall consider operator-valued solutions of the reflection equation. This corresponds to spin chains with boundary degrees of freedom, and is indicated above by having $\bdryk$ act on the additional space $V_B$.

Define the two-row transfer matrix, an endomorphism on $\bigotimes_{i=0}^{L+1} V_i$, as the trace over an auxiliary space $V_A \simeq \C^6$ according to
\begin{equation}
  \mathcal{T}(u) := \tr_A \bigl[\bdryk_{A,L+1}^{t_A}(-u-2) T(u) \bdryk_{A,0}(u) \hat{T}(u) \bigr]\, , \label{eq:Tmatrix}
\end{equation}
where $^{t_A}$ signifies the partial transpose in $V_A$ and we have defined the two monodromies
\begin{equation}
T(u) := \Rmat_{A,L}(u)\Rmat_{A,L-1}(u)\cdots \Rmat_{A,1}(u) \, ,
\end{equation}
\begin{equation}
\hat{T}(u) := \Rmat_{1,A}(u)\Rmat_{2,A}(u)\cdots \Rmat_{L,A}(u) \, .
\end{equation}

By virtue of the Yang-Baxter equation~\eqref{eq:ybe} and reflection equation~\eqref{eq:bybe} the two-row transfer matrix in Eq.~\eqref{eq:Tmatrix} commutes for arbitrary values of the spectral parameter \cite{Sklyanin:1988yz}
\begin{equation}
  [\mathcal{T}(u),\mathcal{T}(v)] = 0 \;, \quad u, v \in \C\, .
\end{equation}

A local open spin chain Hamiltonian is obtained from the transfer matrix according to
\begin{equation}
  \label{eq:H-sim-T-Tinv}
H \sim \mathcal{T}(0)^{-1}\mathcal{T}'(0) \, ,
\end{equation}
where the prime indicates differentiation with respect to $u$, and the precise identification of $H$ requires a choice of normalisation and an additive constant. 
The bulk interactions, $H_{r,r+1}$ for $r = 1, \dots, L-1$ of \eqref{eq:Hform},
depend only on the choice of the $\Rmat$-matrix and are therefore identical to the well-known cyclic case \cite{Minahan:2002ve}. The novel parts are the boundary terms $H_{0,1}$ and $H_{L,L+1}$ that depend on the choice of $\bdryk$-matrices.

The unbroken R-symmetry  $\grp{SU}(2)_H\otimes\grp{SU}(2)_V$ constrains the possible form of $\bdryk$-matrices. Given that the ends should transform trivially under $\grp{SU}(2)_V$, a natural ansatz for the $\bdryk$-matrix is 
\begin{equation}
  [\bdryk(u)]_{i' i} =
  g(u) \tau_{i'}\tau_i+ \tilde{g}(u) \tau_i\tau_{i'} +  f(u) \delta^H_{i'i} + h(u)\delta^V_{i'i} \, , \label{eq:Kansatz}
\end{equation}
where $\delta^{H(V)}$ was defined below  \eqref{eq:end-scale-deriv}. Here $i$ ($i'$) is the
in-going (out-going) index of the auxiliary space, while the $\tau_i$ are matrices acting
on the boundary space.
For $i > 3$ we set $\tau_i=0$,  and for $i=1,2,3$ they form a representation of the the $\alg{su}(2)$ algebra
\begin{equation}
  \label{eq:su-two-alg}
  [\tau_i,\tau_j] := \ii\epsilon_{ijl}\tau_l\,, \quad \epsilon_{123} = 1 \, .
\end{equation}
This ensures that the ansatz Eq.~\eqref{eq:Kansatz} preserves
the $\grp{SU}(2)_H\otimes\grp{SU}(2)_V$ symmetry.
Imposing Eq.~\eqref{eq:bybe} now yields a unique\footnote{It is possible to multiply $\bdryk$
  by an arbitrary function of $u$ without violating \eqref{eq:bybe}, but doing so will only
  contribute a term proportional
  to the identity to \eqref{eq:H-sim-T-Tinv}.}
solution for the undetermined functions, namely
\begin{multline}
  [\bdryk(u)]_{i' i} =
  2u(u+2)\tau_{i'}\tau_i-2u(u+1)\tau_i\tau_{i'}  \\
    - (u+1)(u^2+u+C)\delta^H_{i'i} + (u+1)(u^2+u-C)\delta^V_{i'i} \, . \label{eq:Kmatrix-sol}
\end{multline}
where $C := \sum_{j=1}^3 \tau_j^2$ is the quadratic Casimir.
This solution was found in collaboration with C.~Kristjansen, B.~Pozsgay and 
M.~Wilhelm\cite{kmatToAppear} in the study of integrable matrix product
 states \cite{Piroli:2017sei, Piroli:2018ksf, Pozsgay:2018dzs} 
and overlap formulas for one-point functions \cite{Buhl-Mortensen:2015gfd}.

We find the Hamiltonian from computing the first conserved charge
\begin{multline}
  \label{eq:T-T-inv}
\mathcal{T}(0)^{-1}\mathcal{T}'(0) = 2 + \sum_{r=1}^{L-1} \bigl( 2 P_{r,r+1} + 1 - K_{r,r+1} \bigr) \\ + 2 C^{-1}_{L+1}  \tr_A \bigl[ \bdryk_{A,L+1}^{t_A}(-2) P_{L,A} R_{L,A}'(0) \bigr] - C_0^{-1} \bdryk_{1,0}'(0) \, ,
\end{multline}
where
\begin{equation}
\bigl[ K'(0) \bigr]_{i' i} = 4 \tau_{i'} \tau_i - 2 \tau_i \tau_{i'} - (C+1) \delta_{i' i}^H - (C-1) \delta_{i' i}^V \, .
\end{equation}
Comparing to \eqref{eq:Hform}, \eqref{eq:Hform-bulk} we see that the bulk part matches
if we  identify  
\begin{equation}
  \label{eq:H-from-T}
  H = -\mathcal{T}(0)^{-1}\mathcal{T}'(0) + c
\end{equation}
for some constant $c$.

We shall now see that the dilatation operator for $k=1$ given by Eq.~\eqref{eq:H-k-eq-one} corresponds to an integrable open spin chain Hamiltonian. For $k=1$ the ends of the open-string operators Eq.~\eqref{eq:bdry-op} transform as vectors under $\grp{SU}(2)_H$. To compare we therefore consider the representation $[\tau_j]_{l',l} := - \ii\epsilon_{jl'l} $, for which the left boundary term becomes 
\begin{equation}
 - \frac{1}{2} \bdryk_{0,1}'(0) = \frac{1}{2}  + \mathcal{P}^H_0 (2 P_{0,1} - K_{0,1} ) \mathcal{P}^H_0 \, .
\end{equation}
This is exactly the correct result for our general integrable Hamiltonian to reduce to Eq.~\eqref{eq:H-k-eq-one}! Similarly one shows that $H_{L,L+1}$ is reproduced.

We note that, in this particular case, the boundary terms are given by a projection of the bulk terms onto a subspace. This construction for integrable open spin chains has previously been observed \cite{Nepomechie:2009en,Frahm:1999aa}.

Assuming integrability, the possible form of the dilatation operator is strongly constrained by the symmetries as previously discussed. We can take advantage of this to write down a generalised dilatation operator. For general $k$ the operators in the ends transform in a reducible representation of the R-symmetry; the field $\scalfun$ has two R-symmetry indices, so the boundary sites of our operator \eqref{eq:bdry-op-gen-k} are in
the $\mathbf{3}\otimes \mathbf{k}$ representation of SU$(2)_H$ (but in the trivial SU$(2)_V$ 
representation). The corresponding choice for $\tau$ is then\footnote{Of course $(\scalfun)^\dagger$
  transforms in conjugate representation to that of $\scalfun$, but,
  since we are talking about SU$(2)$,
  the representations will be related by similarity.
  To match field theory exactly, one should thus use
  different, but similar, $t_i$ for the two ends of the spin chain. From the point of view
  of the spectrum, however, one can forget about this detail, since it merely amounts to a change
  of basis of the boundary spaces.}
\begin{equation}
  \label{eq:gen-tau}
  [\tau_j]_{(i',n'),(i,n)} = -\ii \epsilon_{ji'i}\delta_{n'n} + \delta_{i'i}[t_j]_{n',n}\, ,
\end{equation}
where $t_i$ form an irreducible $k$-dimensional representation of \eqref{eq:su-two-alg}.
For $k=1$ we  need to take $t_i$ to be the $1\times 1$ zero matrix in this formula 
(for $k=0$ one should instead take $\tau_i$ to be 
the Pauli $\sigma_i$, as noted in Ref.~\cite{Pozsgay:2018dzs}).
The main claim of this paper is that that the one-loop Hamiltonian for scalar open-string
operators is given by \eqref{eq:T-T-inv} and \eqref{eq:H-from-T}, with $\tau$ as given
above in \eqref{eq:gen-tau} for any $k \geq 1$ .

Let us finally remark that the constant $c$ of \eqref{eq:H-from-T} can be fixed in the usual
way by demanding that chiral primary operators are annihilated. Specifically, there are 
unique boundary states $|\omega_0\rangle$ and $|\omega_{L+1}\rangle$ that have R-charge
$(k+1)/2$ in the 1-2 plane (normalised such that the charge of 
$Z := \phi_1 + \ii \phi_2$ is $+1$). 
Explicitly, we have
\begin{equation}
  (\tau_1+\ii \tau_2)|\omega\rangle = 0\, , \qquad
  \tau_3|\omega\rangle = \tau_{3,\omega}|\omega\rangle \, , \qquad
  \tau_{3,\omega} := \frac{k+1}{2}
\end{equation}
for both $|\omega_0\rangle$ and $|\omega_{L+1}\rangle$.
The identity part of the Hamiltonian is then fixed by demanding
\begin{equation}
  \label{eq:chiral-primary}
  H |\omega_0Z^L\omega_{L+1}\rangle = 0 \, .
\end{equation}

\section{Reflection factors and a consistency check}
\label{sec:reflection}
In this section we subject our proposal to a non-trivial check, by calculating
the asymptotic reflection factors associated with scalar excitations.
To define these, consider an excitation on the BPS vacuum of Eq.~\eqref{eq:chiral-primary}
extended infinitely to the right. The eigenstates take the schematic form
\begin{equation}
  \sum_{r=1}^\infty  \Bigl(  \ee^{-\ii p r} + R(p) \ee^{\ii p r} \Bigr) |r\rangle + 
    \text{boundary terms}\, ,
\end{equation}
where $|r\rangle$ denote the state with the excitation a position $r$, and
$R(p)$ is the reflection factor. In the $k=0$ case, $R(p)$ was calculated for
the two types of scalar excitations in Ref.~\cite{DeWolfe:2004zt}. 
In Refs.~\cite{Correa:2008av,Correa:2011nz} it was further shown that the \emph{ratio} of these factors
is fixed by the supersymmetry preserved by the defect alone, and thus independent of $k$.

We now proceed to determine $R(p)$ for our integrable spin chain. First we consider
a $\phi_4$ excitation ($\phi_5$ and $\phi_6$ are equivalent by the SU$(2)_V$ symmetry).
The ansatz for the eigenstate is 
\begin{equation}
\vert p \rangle_V := \sum_{r=1}^\infty  \Bigl(  \ee^{-\ii p r} + R_V(p) \ee^{\ii p r} \Bigr)\, 
  |\omega_0 Z^{r-1}\phi_4 ZZ\cdots \rangle \, .
\end{equation}
We impose the eigenvalue equation $H \vert p \rangle_V = E_p \vert p \rangle_V$,
with the usual dispersion $E_p = 4 - 2 \ee^{\ii p} - 2 \ee^{-\ii p}$, and find that
\begin{equation}
  R_V(p) = \frac{1}{\ee^{\ii p}} \frac{\tau_{3,\omega} (\ee^{\ii p}-1) + 1}{(\tau_{3,\omega}-1)\ee^{\ii p}-\tau_{3,\omega}} \, .
\end{equation}

The other type of scalar excitation is a $\phi_3$. Since the boundary site is charged
under SU$(2)_H$, it is possible for the $\phi_3$ to mix with a boundary excitation.
Our ansatz is thus
\begin{equation}
  \vert p \rangle_H := \beta \vert \omega_0^-ZZ\cdots\rangle
   + \sum_{r=1}^\infty  \Bigl(  \ee^{-\ii p r} + R_H(p) \ee^{\ii p r} \Bigr)\, 
   \vert \omega_0 Z^{r-1}\phi_3 ZZ\cdots \rangle \, ,
\end{equation}
where  we define $|\omega_0^-\rangle := (\tau_1-\ii \tau_2)|\omega_0\rangle$.
This is an eigenstate for
\begin{equation}
  R_H(p) = -\frac{\tau_{3,\omega}(\ee^{\ii p}-1)+1}{(\tau_{3,\omega}-1)e^{\ii p}-\tau_{3,\omega}}\, ,\qquad
  \beta = \frac{1}{2}\frac{\ee^{\ii p}+1}{(\tau_{3,\omega}-1)\ee^{\ii p}-\tau_{3,\omega}}\, .
\end{equation}
We now observe that $R_V(p)/R_H(p) = -\ee^{-\ii p}$ is indeed independent of $k$ (and in agreement with
the ratio extracted from Ref.~\cite{DeWolfe:2004zt}), even though the two functions have
quite complicated $k$-dependence individually. We take this as a strong indication that our
proposal is correct.

\section{Outlook}

The present work can be extended in several directions. First of all, 
our proposed Hamiltonian should be checked via direct field theory calculations for
$k\geq 2$. We expect this to be straightforward, since the necessary details
of the one-loop perturbation theory is worked out in Ref.~\cite{Buhl-Mortensen:2016jqo}.

In the case of one-point functions of the defect theory it has been possible to match
weak coupling results with string 
theory\cite{Nagasaki:2012re,Buhl-Mortensen:2016pxs,Buhl-Mortensen:2016jqo}
by exploiting a BMN-like limit where one sends $k\to \infty$. It would be interesting to
explore whether something similar is possible for the defect spectrum.

The reflection factors we calculated in the previous section are two of the components
of the reflection matrix. This matrix is, like the S-matrix, fixed by supersymmetry,
up to an overall function\cite{Correa:2008av,Correa:2011nz}. There has been some
progress on determining this overall factor using the crossing equation and
explicit string theory calculations\cite{Correa:2013em}. Hopefully our weak coupling
results can help in constraining it further.

 \vspace{0.2cm}

\section*{Acknowledgements}
We thank Marius de Leeuw, Charlotte Kristjansen, Bal{\'a}zs Pozsgay, Miroslav Rap\v{c}{\'a}k and Matthias Wilhelm for useful discussions. KEV would like to thank Bal{\'a}zs Pozsgay, Charlotte Kristjansen and Matthias Wilhelm for collaboration on a project concerning integrable matrix product states in which the $\bdryk$-matrix Eq.~\eqref{eq:Kmatrix-sol} was found.
ACI would like to thank NBI for kind hospitallity during the completion
of this work.
KEV was supported in part by FNU through grant number DFF-4002-00037,
while ACI was supported by the Villum Foundation.

\vspace*{0.5cm}

\appendix
\section{Renormalisation formulae}
\label{app:renorm}
In position space renormalisation of the UV-divergencies of our
one-loop diagrams amounts to the extension of distributions defined
on $\mathbb{R}^4\setminus\{ 0 \}$ to distributions defined on all of
$\mathbb{R}^4$. A convenient technique is differential renormalisation\cite{Freedman:1991tk}.
The extension of $|z|^{-4}$ is given by the standard formula\cite{Freedman:1991tk}
\begin{equation}
  \label{eq:full-space-mu-deriv}
  \left[\frac{1}{|z|^4}\right]_\mu := -\frac{1}{4}\Box\left(\frac{\log \mu^2|z|^2}{|z|^2}\right)\ ,\qquad
  \mu\frac{\partial}{\partial\mu} \left[\frac{1}{|z|^4}\right]_\mu
    = \frac{1}{8\pi^2}\delta(z)\, .  
\end{equation}
We also need the ``half-space'' version ($\vec z = \{z_0,z_1,z_2\}$)
\begin{equation}
  \label{eq:half-space-extension}
  \left[\theta(z_3)\frac{1}{|z|^4}\right]_\mu := 
  -\frac{1}{4}\Box\left(\theta(z_3)\frac{\log \mu^2|z|^2}{|z|^2}\right)
    + \frac{1}{4}\delta'(z_3)\frac{\log \mu^2|\vec{z}|^2}{|\vec{z}|^2} ,
\end{equation}
\begin{equation}
  \mu\frac{\partial}{\partial\mu} \left[\theta(z_3)\frac{1}{|z|^4}\right]_\mu
     = \frac{1}{16\pi^2}\delta(z) .
  \label{eq:half-space-mu-deriv}
\end{equation}
It is not difficult to check that \eqref{eq:half-space-extension}  is 
indeed an extension. By dimensional 
analysis it then follows that the $\ln \mu$ derivative must be proportional
to $\delta(z)$. The constant of proportionality can be found by integrating
against a suitable  test function, reproducing \eqref{eq:half-space-mu-deriv}.

\section{Boundary fields for $k>1$}
\label{app:bdry-fields}
For $k>1$ the block structure of the fields is still as given in \eqref{eq:block-structure},
but now $\scalsing$ is a $k\times k$ matrix satisfying the singular boundary conditions
\begin{equation}
  \scalsing_i(x) = -\frac{t_i}{x_3}+\text{non-singular}\, ,
  \label{eq:sing-is-sing}
\end{equation}
 where $t_{1,2,3}$ form an irreducible $k$-dimensional representation of the $\alg{su}(2)$ algebra
\eqref{eq:su-two-alg} and $t_{4,5,6} = 0$. 
The index on $\scalfun$ taking values between $k+1$ and $N$ becomes
the fundamental colour index of U$(N-k)$ on the boundary, while the index taking
value from $1$ to $k$ becomes an additional R-symmetry index\cite{deLeeuw:2017dkd}
which we will denote by $n$. The index $n$ transforms in the irreducible $k$-dimensional
representation of SU$(2)_H$.

The singular behaviour \eqref{eq:sing-is-sing} 
of $\scalsing$ makes $\scalfun_i$ go to zero as $x_3^{(k-1)/2}$ for
$i=1,2,3$. We thus define our boundary field as\cite{deLeeuw:2017dkd}
\begin{equation}
  \bscalfun_{i,n}(\vec{x}) := \lim_{x_3\to 0^+}(2x_3)^{-(k-1)/2}
    \left(\scalfun_{i,n}(x)+2\ii\frac{A_{3,n'}^\text{fun}(x) [t_i]_{n',n}}{k+1}\right)\, ,
\end{equation}
for $i=1,2,3$. The $A_3$ term corrects the gauge transformation 
properties\cite{deLeeuw:2017dkd}\footnote{A factor of $2/(k+1)$ is missing
  from (D.8) of Ref.~\cite{deLeeuw:2017dkd}, together with a corresponding factor
  of $1/(\ell+1)$ in (D.7).}.
Since $\scalfun$ has dimension one, the classical dimension of 
$\bscalfun_{i=1,2,3,n}$ is $(k+1)/2$.

Similarly, one can construct boundary fields $\bscalfun_{i=4,5,6}$. Due to the 
different decay properties of $\scalfun_{i=4,5,6}$, these turn out to have
classical dimension $(k+3)/2$.

\section{Connecting $k=0$ with $k > 0$ via partial Higgs\-ing}
\label{app:moving-branes}
The understanding of the $k>0$ defect theory in Ref.~\cite{Gaiotto:2008sa}
is primarily derived from considerations of the moduli space of vacua.
In this appendix we re-derive the basic facts, in the abelian case, using
more pedestrian field theoretic techniques.

In the brane language there is a nice intuitive way to get from $k=0$ to
any $k > 0$:\cite{Gaiotto:2008sa}
 We start from $N$ coinciding D3 branes intersecting a single D5.
We then  take $k$ of D3s on one side of the defect and move them far away
along the D5 (i.e.~along the $\phi_{1,2,3}$ direction). These $k$ (half-)branes 
will decouple, and at low energies we are left with SYM with the rank jumping from 
$N$ to $N-k$
at the D5. 
In the field theory language this construction amount to a partial Higgsing of the $k=0$ theory.
Here we will show how this works for  the abelian case of $N=1$ (and thus $k=1$).\footnote{
  In this $N=1$ case there is complete symmetry between the two sides of the defect. 
  When the separation of the D3 along the D5 is large, the two sides completely decouple. From the
  point of view of one of the D3s the gauge group effectively jumps from U$(1)$ to ``U$(0)$'' (i.e.~the empty theory)
  at the defect.}

The euclidean action for the $k=0$, U$(1)$ theory is\cite{DeWolfe:2001pq}
(we set all Grassman-odd fields to zero for simplicity, suppressed flavour indices are contracted,
and $\hat \mu = 0,1,2$)
\begin{multline}
  S := \int\de^4 x\, \Bigl[\frac{1}{4}F_{\mu\nu}F_{\mu\nu}+\frac{1}{2}(\partial_{\hat\mu}\phi_i)^2
         +\frac{1}{2}(\partial_3 \phi_i + \delta(x_3)q^\dagger \sigma_i q)^2\Bigl] \\
    +\int\de^3 \vec{x}\,\Bigl[(D_{\hat \mu} q)^\dagger D_{\hat\mu}q 
   +q^\dagger q \phi^T \delta^V \phi \Bigr] + S_{\text{g.f.}}\, ,
\end{multline}
where $D_{\hat\mu} q := \partial_{\hat\mu}q - i A_{\hat\mu} q$ and $S_{\text{g.f.}}$
is the gauge fixing. For convenience we extend the Pauli matrices by
 setting $\sigma_{4,5,6} = 0$. Following Ref.~\cite{Gaiotto:2008sa}
we interpret the ill-defined $\delta(x_3)^2$ term as specifying the unusual non-linear boundary 
condition\footnote{
  The derivative of $\phi_i$ satisfying \eqref{eq:bdry-cond} yields a $\delta(x_3)$ term
  exactly cancelling the singular term in the action.}
\begin{equation}
  \Delta\phi_i := \phi_i|_{x_3 = 0^+} - \phi_i|_{x_3 = 0^-} = -q^\dagger \sigma_i q\,  ,
  \label{eq:bdry-cond}
\end{equation}
which turns out to being the key to understanding the fate of the boundary hyper-multiplet.
Looking at small fluctuation around the trivial vacuum, \eqref{eq:bdry-cond} reduces to
$\Delta\phi_i = 0$ at leading order, and we recover the expected free bulk and boundary
spectrum (with e.g.~$S_{\text{g.f.}} = \int\de^4 x \frac{1}{2} \left(\partial_\mu A_\mu\right)^2$).

We now turn to the situation with the D3-brane on the $x_3 < 0$ side 
shifted along the $\phi_{1,2,3}$ direction. This corresponds to setting
\begin{equation}
  \phi_i(x) 
    = \theta(-x_3)\funsc^\dagger \sigma_i \funsc + \scalq_i(x)\, ,\qquad
  q(\vec x) = \funsc+\funsq(\vec x)
\end{equation}
with $\funsc$ independent of $x$ and where $\funsq,\scalq_i$ denote the quantum fluctuations.
When we expand the action to quadratic order around this background we run into 
awkward terms of the form $A_{\hat\mu}\partial_{\hat\mu}\funsq$. This can be cured by the
more exotic  gauge 
\begin{equation}
  S_{\text{g.f.}} = \int\de^4 x \frac{1}{2} \left(\partial_\mu A_\mu
                      -\ii\delta(x_3)[\funsq^\dagger \funsc -\funsc^\dagger\funsq] \right)^2 .
\end{equation}
Here we again encounter a $\delta(x_3)^2$ term, which we translate to the 
boundary condition
\begin{equation}
  \Delta A_3 := A_3|_{x_3 = 0^+} - A_3|_{x_3 = 0^-} = \ii[\funsq^\dagger \funsc -\funsc^\dagger\funsq]\, .
  \label{eq:bdry-cond-A}
\end{equation}
The fact that we get an additional boundary condition solves another problem
for us; the field $q$ has 
four real components, so \eqref{eq:bdry-cond-A} together with the linear truncation
of \eqref{eq:bdry-cond}, $\Delta\scalq_i = -\funsc^\dagger\sigma_i\funsq-\funsq^\dagger\sigma_i\funsc$,
provides exactly the right number of equations to solve for $q$ in terms of $\Delta\phi_{1,2,3}$
and $\Delta A_3$.
Doing this we arrive at the following quadratic theory
\begin{multline}
  S_0 =  
    \int\de^3 \vec{x}\,\Bigl[\frac{1}{4\funsc^\dagger\funsc}[(\partial_{\hat\mu}\Delta A_3)^2
       +(\partial_{\hat\mu}\Delta\scalq)^T\delta^H (\partial_{\hat\mu}\Delta\scalq) ] \\
   +\funsc^\dagger \funsc (A_{\hat\mu}A_{\hat\mu}+\phi^T \delta^V \phi) \Bigr]
   + \int\limits_{\mathbb{R}^4\setminus \text{def.}}\de^4 x \,
       \Bigl[\frac{1}{2}(\partial_\mu A_\nu)^2+\frac{1}{2}(\partial_{\mu}\scalq_i)^2\Bigl]
  \label{eq:S-free}
\end{multline}
only involving bulk fields.\footnote{Since some of the fields are discontinuous, 
  the meaning of the bulk integration need to be specified. We  set
  $
    \int_{\mathbb{R}^4\setminus \text{def.}}\de^4 x := \int\de^3 \vec{x}
              \left(\int_{-\infty}^{0^-} \de x_3 + \int_{0^+}^{\infty}\de x_3\right).
  $}

When the separation between the D3s on each
side is large (i.e.~when $\funsc^\dagger\funsc$ is much larger than the energy scale 
of the excitation), we can neglect the first term of \eqref{eq:S-free}. This means
that the boundary conditions for the hyper-multiplet ($\scalq_{1,2,3}$,$A_3$)
are `free', i.e.~Neumann. On the other hand, the localised mass-like term for the 
vector-multiplet ($\phi_{4,5,6}$,$A_{\hat\mu}$) is very large, leading to Dirichlet
boundary conditions. We have thus demonstrated, for $k=1$, both that there is no independent
defect hyper-multiplet, and that  the explicit boundary conditions given in Ref.~\cite{Gaiotto:2008sa} emerge.

\end{document}